\documentclass{IEEE_lsens}
\usepackage{textcomp}

\usepackage{xcolor}

\usepackage[noadjust]{cite}

\ifCLASSINFOpdf
   \usepackage[pdftex]{graphicx}
   \graphicspath{{../pdf/}{../jpeg/}}
   \DeclareGraphicsExtensions{.pdf,.jpeg,.png}
\else
\fi

\usepackage[T1]{fontenc} 
\usepackage{amsmath}
\usepackage[cmintegrals]{newtxmath}
\usepackage{bm}

\usepackage{array}

\usepackage[caption=false,font=footnotesize,labelfont=sf,textfont=sf]{subfig}
\usepackage{amssymb}
\usepackage{tikz}

\usepackage{url}
\ifCLASSINFOpdf
\else
\fi
\providecommand{\hypersetup}[1]{\relax}

\hypersetup{pdftitle={Measurement Series for Detection of Critical Points in 600 Scanning Micro Mirrors Affected by Three-Wave Down-Conversion},
pdfsubject={Sensor Phenomena: Modeling and Simulations},
pdfauthor={Ulrike Nabholz},
pdfkeywords={nonlinear dynamics, mems, resonator, mode coupling, critical point, modeling,  down-conversion}}
\newcommand\copyrighttext{%
	\footnotesize \textcopyright 2020 IEEE.Personal use of this material is permitted. Permission from IEEE must be obtained for all other uses, in any current or future media, including reprinting/republishing this material for advertising or promotional purposes,creating new collective works, for resale or redistribution to servers or lists, or reuse of any copyrighted component of this work in other works.}
\newcommand\copyrightnotice{%
	\begin{tikzpicture}[remember picture,overlay]
	\node[anchor=south,yshift=10pt] at (current page.south) {\fbox{\parbox{\dimexpr\textwidth-\fboxsep-\fboxrule\relax}{\copyrighttext}}};
	\end{tikzpicture}%
}

\begin{document}



%
%
\title{Validating the Critical Point of Spontaneous Parametric Down-Conversion for over 600 Scanning MEMS Micro Mirrors on Wafer-Level}

%
\author{\IEEEauthorblockN{Ulrike~Nabholz\IEEEauthorrefmark{1}, Florian~Stockmar\IEEEauthorrefmark{2}, Jan~E.~Mehner\IEEEauthorrefmark{3}
and~Peter~Degenfeld-Schonburg\IEEEauthorrefmark{1}}
\IEEEauthorblockA{\IEEEauthorrefmark{1}Robert Bosch GmbH, Corporate Research, 71272 Renningen, Germany\\
\IEEEauthorrefmark{2}Robert Bosch GmbH, Automotive Electronics, 72762 Reutlingen, Germany\\
\IEEEauthorrefmark{3}Technical University of Chemnitz, 09107 Chemnitz, Germany\\
}%
\thanks{Corresponding authors: U. Nabholz (e-mail: ulrike.nabholz@de.bosch.com), P. Degenfeld-Schonburg (e-mail: peter.degenfeld-schonburg@de.bosch.com).\protect}
}
%
%
%


\IEEEtitleabstractindextext{%
\begin{abstract}
Sensors and actuators based on resonant micro-electro-mechanical systems (MEMS), such as scanning micro mirrors, are well-established in automotive and consumer products. As the areas of application broaden towards highly-automated driving and augmented reality, the performance requirements for the MEMS are also increasing. Devices outside of the performance specifications have to be rejected which is costly due to the high processing times of MEMS technologies. In particular, nonlinear system behavior is often found to cause unexpected device failure or performance issues. Thus, accurate simulation or rather system models which account for nonlinear sensor dynamics can not only increase process yield, but more importantly, lead to a comprehensive understanding of the underlying physics and consequently to improved MEMS design. In a recent work \cite{Nabholz2019a}, we have studied the possibility of a rather drastic device failure induced by nonlinearities on the example of a resonant scanning MEMS micro mirror. On the level of a few selected chips, we have carefully measured the complex nonlinear system behavior and modelled it by a nonlinear mode-coupling phenomenon known as spontaneous parametric down-conversion (SPDC). The most intriguing feature of SPDC is the sudden change from a rather linear to a highly nonlinear system behavior at the threshold or rather critical oscillation amplitude of the mirror. However, the threshold only lies within the range of the mirror's operational amplitude, if certain frequency resonance conditions regarding the modes of the mechanical structure are met. As a direct consequence, the critical amplitude strongly depends on the frequency spectrum of the MEMS design which in turn is largely influenced by fabrication imperfections. In this work, we validate the dependence of the critical amplitude on the resonance condition by measuring it for over 600 micro mirrors on wafer-level. Our work does not only validate the theory of SPDC with measurements on such a large scale, but also demonstrates modeling strategies which are essential for MEMS product design. 		
\end{abstract}

}


\maketitle

\copyrightnotice

\section{Introduction}

\begin{figure}[h]
	\centering
	\includegraphics[width=0.85\linewidth]{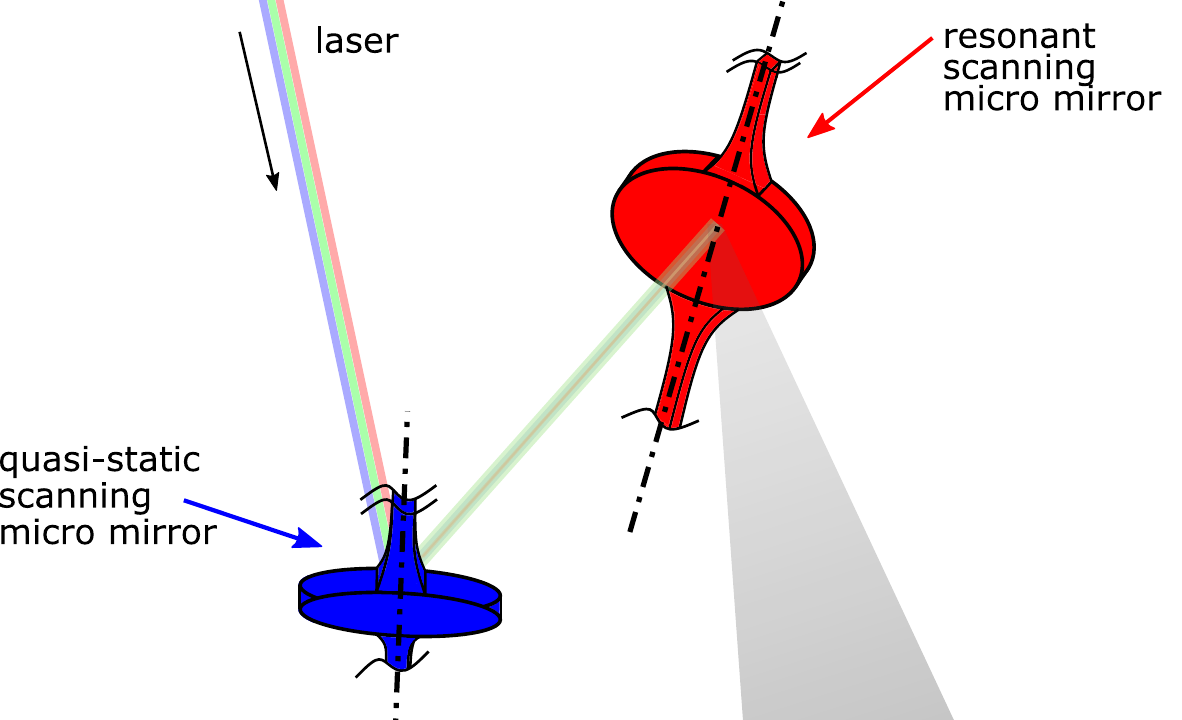}
	\caption{Based on \cite{BST2017}: Working principle of the BML050 microscanner. The red structure denotes the resonantly operated micro mirror that will be analysed in the following. For the sake of completeness, the blue structure performs the quasi-static oscillation and is thus not of interest here. The dotted-dashed lines indicate the resonant and the quasi-static axes of oscillations.}
	\label{fig:working_principle}
\end{figure}

Scanning micro mirrors constitute key devices for applications in the automotive as well as the consumer electronics sector. Designed as resonant micro-electro-mechanical systems (MEMS), they enable laser distance measurements as well as image projection for augmented reality applications where miniaturization is crucial \cite{Solgaard2014, Yalcinkaya2006}.  Within the field of micro-opto-electro-mechanical systems (MOEMS), the first devices to emerge were Digital Micromirror Devices \cite{Mignardi2000}, where each mirror in an array represents one pixel that can be switched between different states. In contrast, today's scanning micro mirrors are characterized by resonant operation of a torsional degree of freedom and need only a single mirror for projection \cite{Petersen1980, Baran2012, Kurth1998}: A laser beam is pointed onto the mirror's reflective structure which oscillates with a resonance frequency in the kHz range. Overall, the design typically aims for the linear regime, meaning that during the device operation only a single mode is supposed to be excited by the external actuation. The working principle is shown in Fig. \ref{fig:working_principle}. 
For small deflection angles, the mirror's oscillation can be approximated by a linear model. At higher deflection angles, nonlinear terms become relevant and can limit the maximum deflection angle, as will be shown here.
\subsection{Mode Coupling Phenomenology}
Mechanical mode coupling phenomena in scanning micro mirrors are known \cite{Baskaran2003} and the principle of three-wave mixing has already been observed \cite{Ganesan2016} and modelled \cite{Ganesan2017} for simpler MEMS geometries. 
As introduced in our previous work \cite{Nabholz2019a}, nonlinear mode coupling in scanning micro mirrors, i.e. the actuation of unwanted ('parasitic') modes, reduces the attainable drive amplitude and can even lead to fracture of the mechanical structure (in places, where the deflection caused by parasitic modes induces high mechanical stress). We adapted a process known from the field of nonlinear optics as spontaneous parametric down-conversion (SPDC) by three-wave mixing \cite{Boyd2008, Carmichael2009, Degenfeld-Schonburg2015} to the nonlinear dynamics of the MEMS micro mirror originating from structural or rather geometric nonlinearities \cite{Nayfeh2008, Strogatz2007, Mestrom2008, Kacem2009, Lifshitz2009}. \\
On the level of single chips we have reported on experimental observations showing a variety of nonlinear dynamical behavior ranging from stationary state bifurcations to dynamical instabilities. Using a three degree-of-freedom model, with the relevant parameters extracted from the experiments, we have been able to model all the observed phenomena with high accuracy. \\
Most importantly, SPDC is a critical phenomenon, i.e. disturbances only occur above a certain threshold deflection (or critical point) of the drive mode. Below threshold, a single degree-of-freedom nonlinearly damped Duffing oscillator provides an accurate estimate of the device behaviour \cite{Duffing1918, Nabholz2018}. \\
Here, in accordance with SPDC, we can observe resonant actuation of two parasitic modes, denoted by the indices \(1\) and \(2\), in some devices, whenever the linear mode frequencies \(f_{0,1}\) and \(f_{0,2}\) fulfil the condition \(f_{{0,0}} \approx f_{{0,1}} + f_{{0,2}}\) (with the drive mode linear frequency denoted by \(f_{0,0}\)). Fig. \ref{fig:3dof} provides an overview of the possible observations which will be discussed below.
In addition to the physical insights, the most important quantity for MEMS product design is the critical amplitude (proportional to the deflection angle) or rather critical point of SPDC. The critical point marks the onset of the parasitic mode excitations and depends only on the three modes of oscillation that are relevant for the three-wave mixing in question. Despite the complexity of the nonlinear model, it can be expressed analytically which allows to deduce measures for MEMS design to prevent SPDC within the desired operational ranges. \\
In this work, we show how an extraction of the relevant linear and nonlinear parameters from a single chip measurement allows us to apply the critical point model onto all chips of the same design. 
The so-called 'design parameter' which describes the frequency mismatch of the three modes, is given by
\begin{equation}
\delta = \omega_{0,0} - \left(\omega_{0,1} + \omega_{0,2}\right)
\end{equation}
and dominates the critical point over all chips on the wafer. In Section \ref{sc:test}, we measure both the resonance conditions, which are a simple combination of linear mode frequency, as well as the critical point for 631 chips on wafer-level to prove the accuracy of the critical point model.   
\section{Mathematical Model}
We provide an overview of the mathematical model derived in our previous work \cite{Nabholz2019a} and highlight the importance of the critical point calculation.
For completeness, we state the modal form \cite{Kim2014} of the three coupled equations of motion which are given by 
\begin{IEEEeqnarray}{ll}
	\ddot{q}_0 + \frac{\omega_{0,0}}{Q_0}\dot{q}_0 + \omega_{0,0}^2 q_0 + \tilde{\beta_{0}} q_0^3 + \frac{\omega_{0,0}}{Q_{\mathrm{nl},0}}q_0^{2} \dot{q_0} & + \tilde{\alpha} q_1 q_2 \nonumber\\
	& = F_\mathrm{0} \sin\left(\omega_{\mathrm{osc,0}} t\right), \label{eq:3dof_ode0} \\
	\vspace{10pt}
	\ddot{q}_1 + \frac{\omega_{0,1}}{Q_1}\dot{q}_1 + \omega_{0,1}^2 q_1 + \tilde{\beta_{1}} q_1^3 + \frac{\omega_{0,1}}{Q_{\mathrm{nl},1}}q_1^{2} \dot{q_1} & + \tilde{\alpha} q_0 q_2 = 0, \label{eq:3dof_ode1}\\
	\ddot{q}_2 + \frac{\omega_{0,2}}{Q_2}\dot{q}_2 + \omega_{0,2}^2 q_2 + \tilde{\beta_{2}} q_2^3 + \frac{\omega_{0,2}}{Q_{\mathrm{nl},2}}q_2^{2} \dot{q_2} & + \tilde{\alpha} q_0 q_1 = 0. \label{eq:3dof_ode2}
\end{IEEEeqnarray}
Table \ref{table} shows the parameters used in the above equations as well as in the critical point model later on in equation \eqref{eq:crit_pll}. The variables of the system are given by the time-dependent modal amplitudes \(q_0\), \(q_1\) and \(q_2\) for the modal amplitudes of the drive mode, parasitic mode \(1\), and parasitic mode \(2\), respectively. The dot represents the time-derivative and \(\omega_{\mathrm{osc,0}}\) denotes the oscillation frequency of the drive mode which is identical to the external drive frequency. The tilde operator is used to simplify the notations \(\beta_{n} = \frac{3 \tilde{\beta}_n}{\omega_{\mathrm{0,0}}^2}\) and \(\alpha = \frac{\tilde{\alpha}}{2\sqrt{2 \omega_{\mathrm{0,0}} \omega_{\mathrm{0,1}} \omega_{\mathrm{0,2}}}}\). 
\begin{table}[h]
	\begin{tabular}{l|c}
		\hline
		\textbf{Parameter} & \textbf{Symbol (Mode index n = 0, 1, 2)} \\
		\hline \hline
		Linear mode frequency & \(f_{\mathrm{0},n}\) \\
		\hline
		Angular mode frequency & \(\omega_{\mathrm{0},n} =  2\pi f_{\mathrm{0},n}\) \\
		\hline
		Angular oscillation frequency & \(\omega_{\mathrm{osc},n}\)\\
		\hline
		Input force amplitude & \(F_{\mathrm{0}}\) \\
		\hline
		Linear quality factor & \(Q_n\) \\
	    \hline
		Nonlinear quality factor & \(Q_{\mathrm{nl},n}\) \\
		\hline
		Duffing coefficient & \(\beta_n \propto \tilde{\beta}_n\)\\
		\hline
		Three-wave coupling coefficient & \(\alpha  \propto \tilde{\alpha}\) \\
		\hline
		\end{tabular}
	\vspace{5pt}
	\caption{Parameters in equations \eqref{eq:3dof_ode0}-\eqref{eq:3dof_ode2}.}
	\label{table}
\end{table}
\noindent
In \cite{Nabholz2019a}, we have analysed equations \eqref{eq:3dof_ode0}-\eqref{eq:3dof_ode2} in detail which is beneficial for understanding the full range of nonlinear dynamics and for the extraction of the system parameters. Yet, for predicting the onset of SPDC, the critical point model suffices. To grasp the concept of a critical point, without going through the details of the analysis, we first note that equations \eqref{eq:3dof_ode0}-\eqref{eq:3dof_ode2} can be solved by the trivial solution \(q_1 = q_2 = 0\) for every point in time. In this case, the drive mode oscillates according to a nonlinearly damped and driven Duffing oscillator \cite{Nabholz2018}. However, at the critical point of the drive mode's stationary state oscillation amplitude (which denotes the onset of SPDC) corresponding to a critical deflection angle, this trivial solution becomes unstable in favour of a stable solution with \(q_1 \neq 0, q_2 \neq 0\). \\
The critical amplitude \(a_0^{\mathrm{crit}}\) can be derived from the steady-state solutions of equations \eqref{eq:3dof_ode0}-\eqref{eq:3dof_ode2} obtained using averaging methods \cite{Nabholz2019a} with the relation \(q_0 = a_0 \sin\left(\omega_{\mathrm{osc},0} t\right)\). It needs to be calculated individually for each device, since process tolerances during MEMS fabrication influence the mode spectrum and thus, the scope of possible mode couplings. \\
When the phase difference between the actuation force and the drive mode response is controlled by a phase-locked-loop (PLL) \cite{Lee2011, Nabholz2018} to ensure an actuation at the resonance frequency of the drive mode, the critical amplitude is given by
\begin{align}
\label{eq:crit_pll}
\left(a_{0,\mathrm{PLL}}^{\mathrm{crit}}\right)^2  = & \frac{-2 d_{\mathrm{1}} d_{\mathrm{2}} \beta_0 \delta +  \left(d_{\mathrm{s}} \alpha^2 \pm \sqrt{d_{\mathrm{s}}^2 \alpha^4 - 4 d_{\mathrm{1}} d_{\mathrm{2}} \beta_0 \left(d_{\mathrm{1}} d_{\mathrm{2}} \beta_0 + \alpha^2 \delta\right)}\right)}{2 d_{\mathrm{1}} d_{\mathrm{2}} \beta_0^2} 
\end{align}
To improve readability, the damping terms \(d_{\mathrm{1}} = \frac{\omega_{\mathrm{0,1}}}{2 Q_1}, d_{\mathrm{2}} = \frac{\omega_{\mathrm{0,2}}}{2 Q_2}\) and the short notation for the sum of the two damping terms are used: \(d_\mathrm{s} :=\frac{\omega_{\mathrm{0,1}}}{2 Q_{\mathrm{1}}} +  \frac{\omega_{\mathrm{0,2}}}{2 Q_{\mathrm{2}}} \). 
The critical deflection angle is independent of the nonlinear damping and the Duffing terms of the parasitic modes, since the two parasitic modes only gain amplitude above the threshold, thus rendering their nonlinear damping and Duffing term irrelevant. Between devices of the same design, \(\delta\) is the only parameter that varies significantly.\\
For the special case of small Duffing coefficients in the drive mode, as observed in the tested micro mirror, the simplified form of equation \eqref{eq:crit_pll} is given by
\begin{equation}
\left(a_{0,\mathrm{PLL}}^{\mathrm{crit}}\right)^2 \xrightarrow{\beta_0 \rightarrow 0} \frac{d_{\mathrm{1}} d_{\mathrm{2}}}{\alpha^2}\left(1 + \frac{\delta^2}{d_{\mathrm{s}}^2}\right). \label{eq:crit_pll_short}
\end{equation}
Let us now comment on the most important insight that can be drawn from equation \eqref{eq:crit_pll_short} (and also from equation \eqref{eq:crit_pll} for the more general case). The critical amplitude depends on design parameter \(\delta\), damping coefficients \(d_1, d_2\), three-wave coupling coefficient \(\alpha\), linear resonance frequencies \(\omega_{0,0}\), \(\omega_{0,1}\), \(\omega_{0,2}\) and drive mode Duffing coefficient \(\beta_0\) of the devices. In principle all of these parameters will differ from device to device within the range of a few percent due to differences in the geometry as a consequence of fabrication imperfections. However, these small changes cause changes of several orders of magnitude for \(\delta^2\) which describes the resonance condition. Thus, the changes in \(\delta\) from device to device will dominate the outcome of the critical point above all other deviations between devices of the same design. Therefore, we propose to extract the quantities for the coefficients from a single chip measurement as detailed in \cite{Nabholz2019a} and deduce the critical point for all chips from the simple functional dependence on \(\delta\) as given by equations \eqref{eq:crit_pll} and \eqref{eq:crit_pll_short}.    
\section{Measurements}
\label{sc:test}
\noindent Scanning micro mirrors are designed to reach an application-specific deflection angle \(\phi \propto a_0\) up to a maximum of \(\phi = \phi_{\mathrm{limit}}\), for which an approximate value of \(9^{\circ}\) in each direction is specified in \cite{BST2017}. \\
\begin{figure}[h]
	\centering
	\includegraphics[width=1.0\linewidth]{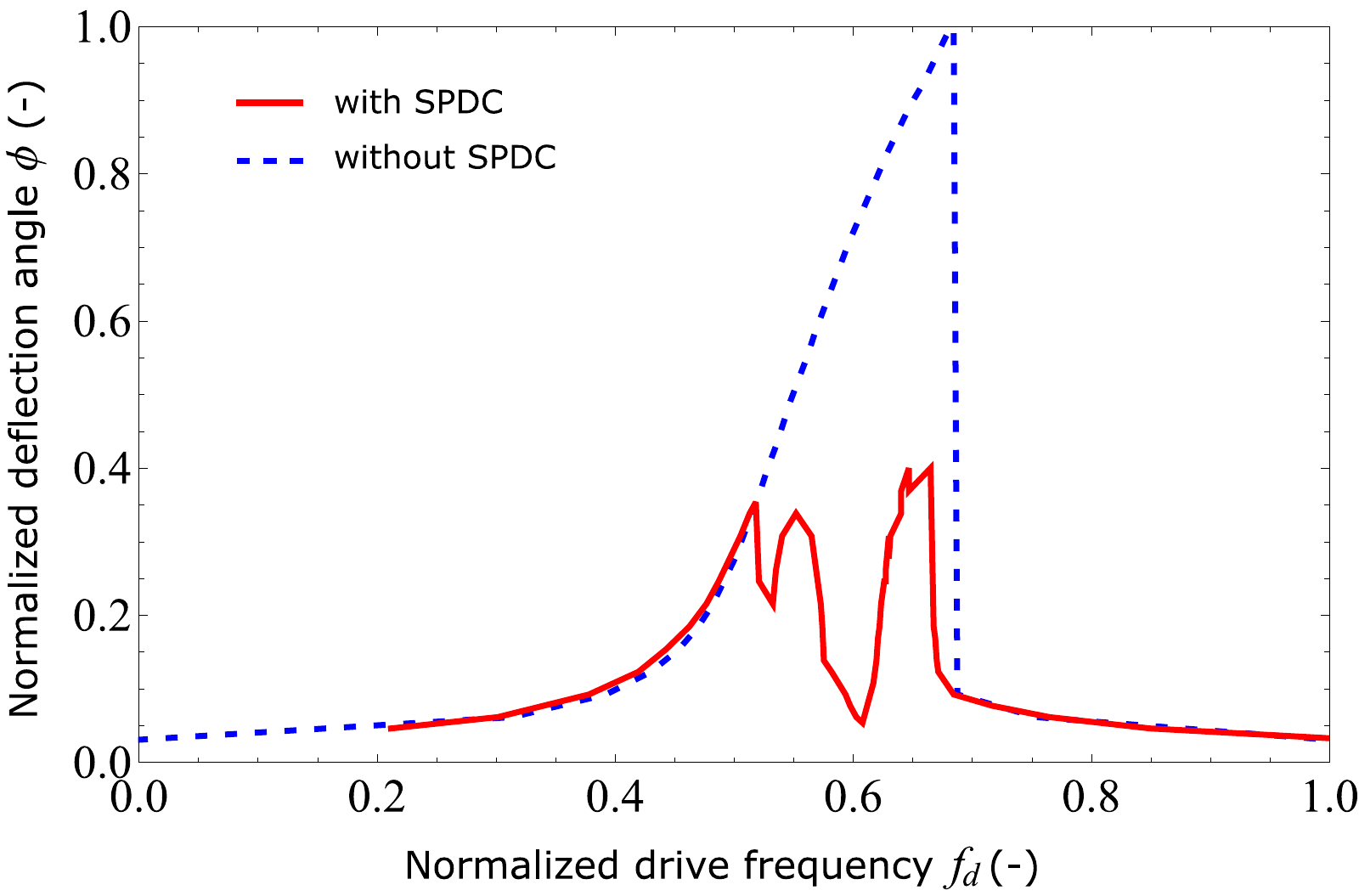}
	\caption{Optically measured amplitude response curves for two micro mirrors with different values of \(\delta\). One displays SPDC (red, solid), the other does not (blue, dashed). Note that the frequency is normalized to include only the measured range for each case in order to ensure comparability and that this geometry is slightly different from the one analysed in Fig. \ref{fig:results}, mainly leading to \(\beta_0 \neq 0\). Mode coupling effects occur, starting at the critical point where the two curves diverge. }
	\label{fig:3dof}
\end{figure}
In contrast to the high measurement effort needed to extract the set of 13 parameters which characterize the full range of dynamic effects due to SPDC that are shown in Fig. \ref{fig:3dof}, for the critical point model the extraction of only seven parameters suffices. Table \ref{table2} shows the full parameter set and its classification.
	\begin{table}[h]
	\begin{tabular}{c|c|c}
		\hline 
		\multicolumn{3}{c}{\textbf{Parameter}}\\
		\hline \hline
		\(f_{\mathrm{0,0}}\) & \(f_{\mathrm{0,1}}\) & \(f_{\mathrm{0,2}}\) \\
		\hline
		\multicolumn{3}{c}{\(\mathbf{\alpha}\)} \\
		\hline
		\(\beta_0\) & \color{gray}{\(\beta_1\)} & \color{gray}{\(\beta_2\)}  \\
		\hline
		\color{gray}{\(Q_{\mathrm{nl},0}\)} & \color{gray}{\(Q_{\mathrm{nl},1}\)} & \color{gray}{\(Q_{\mathrm{nl},2}\)} \\
		\hline
		\color{gray}{\(Q_0\)} & \(Q_1\) & \(Q_2\)\\
		\hline
	\end{tabular}
	\vspace{5pt}
	\caption{Parameters. Black: relevant to critical point model; grey: only relevant to full system model.}
	\label{table2}
\end{table}
\noindent
Apart from the coupling coefficient \(\alpha\) and the Duffing coefficient \(\beta_\mathrm{0}\), all relevant parameters for the critical point model (shown in black in Table \ref{table2}) can be directly measured using standard MEMS characterization tests. The Duffing coefficient \(\beta_\mathrm{0}\) is obtained by performing closed-loop measurements of a single device below the critical point and tracking the frequency shift as the amplitude of the actuation force is increased \cite{Nabholz2018, Polunin2016}. 
Once the frequency and Duffing parameter of the drive mode in addition to the quality factors and frequencies of the parasitic modes are measured for the single chip the three-wave coupling \(\alpha\) can be deduced from the measurement of the critical point and equation \eqref{eq:crit_pll} \cite{Nabholz2019a}. Once again, we emphasize that a single chip measurement suffices. \\
For the large-scale evaluation of the critical point, we have measured a set of 631 devices. Under a PLL-controlled actuation we have increased the actuation force step by step. Each level is held for \(5\,\mathrm{s}\) in order to allow for a long enough waiting period until the onset of SPDC. The actuation force is increased, as long as no disturbance in the electrical signal is detected. As soon as disturbances occur (e.g. due to a beating in the time signal originating from the actuation of the two parasitic modes oscillating at different frequencies), the current deflection angle is documented as the critical deflection angle \(\phi\). In addition, we have measured the resonance frequencies of the mode triplet and thus the parameter \(\delta\) for each mirror. 
\section{Results}
\begin{figure}[b]
	\centering
	\includegraphics[width=1.0\linewidth]{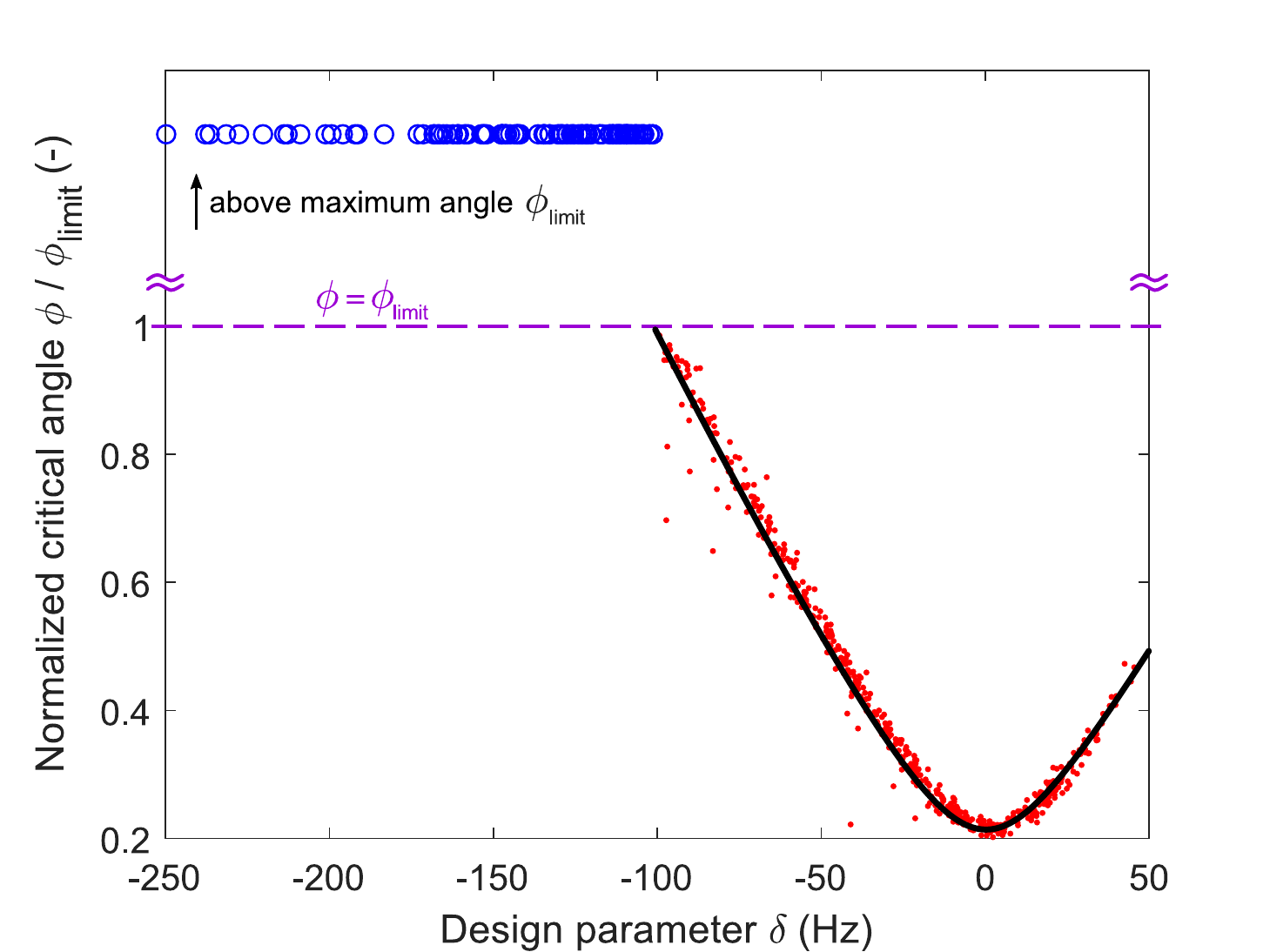}
	\caption{Measured critical deflection angles for the onset of mode coupling compared to the prediction from theory, with red dots and blue circles denoting parts with critical angles below and above the specified maximum angle \(\phi_\mathrm{limit}\), respectively. The predicted threshold value curve is provided by the black line.}
	\label{fig:results}
\end{figure}
\noindent Fig. \ref{fig:results} shows the comparison of the measured critical deflection angle \(\phi\) with red dots and blue circles denoting parts with critical angles below and above the specified maximum angle \(\phi_\mathrm{limit}\), respectively, with the predicted threshold value from the model for different design parameters \(\delta\) shown by the black solid line. We emphasize that the black line is calculated with parameters for frequencies, quality factors, and three-wave coupling coefficient \(\alpha\) as extracted from a single-chip measurement.  \\
\noindent The results show that our prediction agrees very well with the vast majority of measurements within the evaluated parameter range. A few outliers occur that exhibit signs of mode coupling already at a lower deflection angle than predicted. Yet, no instance is recorded, where mode coupling was predicted for a certain angle, but occurs much later or not at all. This validates our hypothesis that SPDC occurs in these devices and that the critical point model can be used to forecast the threshold deflection angle of all chips after extraction of the system parameters from a single chip measurement only. \\
The threshold angle is sensitive even to small changes in the linear mode spectrum of a mirror device which shifts slightly due to geometric differences induced by MEMS fabrication imperfections. Thus, it is not only design-specific, but highly device-specific for devices of the same design. \\
The comparison of measurements with simulations confirms our model and reveals that even only a few Hz change in the resonance criterion (given by the design parameter \(\delta\)) can decide about the occurrence of mode coupling.
\section{Conclusion}
\noindent By performing critical point measurements on 631 parts of the same design, we showed the validity of our model for the relevant operational state of a scanning micro mirror. Since the design parameter \(\delta\) provides a handle for tuning the mode coupling behaviour of a design, the relevance of the critical deflection angle model to MEMS design in the product development phase is also highlighted: Changing the frequency of one or more of the parasitic modes by a few hundred \,Hz strongly influences the design parameter \(\delta\) and can thus alleviate SPDC entirely.\\
This provides an efficient and reliable way of identifying unwanted nonlinear mode coupling effects. If applied early on in the MEMS development process, it can contribute significantly towards reducing the number of design iterations needed by providing a means of checking a given design layout for possible couplings and re-iterating the simulated design if necessary. \\
Our model for the critical amplitude given in equation \eqref{eq:crit_pll_short} can be applied to any oscillatory MEMS sensor or actuator with a fitting frequency match that also fulfils the conditions of separable fast and slow time scales as well as high quality factors. 
Here, similar coupling mechanisms for resonant systems, such as four-wave mixing or three-wave mixing effects of different origins are also conceivable \cite{Nabholz2019b}. Due to the versatility of the developed method, it can also be applied to high-Q oscillatory systems outside of the MEMS domain.
\normalsize
\bibliographystyle{IEEEtran}

\bibliography{IEEEabrv,basic_bib}
%
%
%
%

\end{document}